\documentclass{aa}
\usepackage{graphicx}
\usepackage{natbib}

\begin{document}
   \title{Multiple outflows in IRAS 19410+2336
\thanks{Based on observations with the IRAM Plateau de Bure 
Interferometer, the Calar Alto 3.5\,m telescope and the IRAM
30\,m. IRAM is supported by INSU/CNRS (France), MPG (Germany), and IGN
(Spain). Calar Alto is operated by the MPIA in Heidelberg, jointly
with the Spanish National Commission for Astronomy.}}
   \titlerunning{Multiple outflows in IRAS 19410+2336}
   \authorrunning{Beuther et al.}

   \author{H. Beuther\inst{1,2}, P. Schilke\inst{1} and T. Stanke\inst{1}}
   \offprints{H. Beuther}    
   \institute{Max-Planck-Institut f\"ur Radioastronomie, Auf dem H\"ugel 69, 53121 Bonn, Germany \and Harvard-Smithsonian Center for Astrophysics, 60 Garden Street, Cambridge MA 02138, USA\\ \email{beuther@mpifr-bonn.mpg.de,hbeuther@cfa.harvard.edu}}

\date{Received March 24th 2003; accepted May 26th 2003}

   \abstract{Plateau de Bure Interferometer high-spatial resolution CO
   observations combined with near-infrared H$_2$ data disentangle at
   least seven (maybe even nine) molecular outflows in the massive
   star-forming region IRAS\,19410+2336. Position-velocity diagrams of
   the outflows reveal Hubble-like relationships similar to outflows
   driven by low-mass objects. Estimated accretion rates are of
   the order $10^{-4}$\,M$_{\odot}$\,yr$^{-1}$, sufficiently high to
   overcome the radiation pressure and form massive stars via
   disk-mediated accretion processes. The single-dish large-scale mm
   continuum cores fragment into several compact condensations at the
   higher spatial resolution of the PdBI which is expected due to the
   clustering in massive star formation. While single-dish data give a
   simplified picture of the source, sufficiently high spatial
   resolution resolves the structures into outflows resembling those
   of low-mass star-forming cores. We interpret this as further
   support for the hypothesis that massive stars do form via
   disk-accretion processes similar to low-mass stars.
 }

   \maketitle
%

\section{Introduction}

The physical processes forming massive stars are still subject to
great debate. Basic low-mass star-forming theories predict small
accretion rates independent of mass
($10^{-6}-10^{-5}\,$M$_{\odot}$yr$^{-1}$, \citealt{shu 1977}), which
are not sufficient to overcome the radiation pressure of a star $\geq
8$\,M$_{\odot}$. More massive stars should not form according to that
picture \citep{wolfire 1987} although they are observed to
exist. Several theories have been put forward to solve this
discrepancy: if the standard scenario is adapted, higher accretion
rates must be present and the accretion is likely to be mediated
through disks (e.g., \citealt{wolfire 1987,jijina 1996,norberg
2000,maeder 2002,yorke 2002,mckee 2002}). Another scenario suggests
that the very center of the evolving cluster is dense enough that
intermediate-mass protostars collide, merge, and form the most massive
objects via these collisions (e.g.,
\citealt{bonnell 1998,stahler 2000,bally 2002,zinnecker 2002}).

As high-mass star-forming regions are on the average far away (a few
kpc) and massive stars form in a clustered mode, it is nearly
impossible to resolve with current telescopes the forming cluster in
the mm and sub-mm regime, and thus to study the physical processes
forming the stars in a direct way. Infrared observations give ambiguous
results at best, since the centers of activity are very deeply
embedded and block out all infrared light, except in rare favorable
geometries. Therefore, indirect approaches have to be taken. 
Massive molecular outflows are perfectly suited for such
investigations, because outflows are observed on large spatial scales
(in the parsec regime), they are easier to resolve spatially, and
their structure and energetics give information on the physical
processes close to their driving center. For example, the accretion
scenario predicts outflows which are morphologically similar to
low-mass outflows and which show a high degree of collimation due to
the star-disk interaction \citep{richer 2000}. On the contrary,
colliding protostars are expected to be extremely eruptive phenomena
during which accretion disks should not be able to survive, and
therefore the outflows are likely less collimated and rather resemble
explosions as observed in Orion KL \citep{schultz 1999}.

Over recent years several single-dish studies of massive outflows have
been undertaken (e.g., \citealt{shepherd 1996,ridge 2001,zhang
2001,beuther 2002b}). While these studies agree on the facts that
outflows are ubiquitous phenomena in massive star formation and that
they are very massive and energetic, there is considerable
disagreement on the typical collimation of the observed
outflows. \citet{shepherd 1996} and \citet{ridge 2001} claim that the
average collimation of massive outflows is lower than observed for
their low-mass counterparts supporting the proposal that massive star
formation has to proceed through different physical processes such as
coalescence. \citet{beuther 2002b} also observed lower collimation of
massive outflows compared to low-mass sources. However, taking
properly into account the poor spatial resolution, they argue that
their data are well consistent with highly collimated outflows in
high-mass star-forming regions (but proof of this requires
interferometer observations). Thus, no other physical processes
have to be invoked, and high-mass star formation can proceed as in the
classical low-mass scenario, just with significantly enhanced
accretion rates.

Despite the great advances during recent years, single-dish
observations are not sufficient to understand the complex outflows in
massive star formation. As always clusters of stars form
simultaneously, it is likely that at high spatial resolution many of
the single-dish outflows resolve into multiple outflows emanating from
the same large scale core but from different protostars within them.
Using Plateau de Bure interferometric observations \citet{beuther
2002d} showed that the chaotic single-dish outflow in IRAS\,05358+3543
resolves into at least three molecular outflows at a spatial
resolution of $\sim 3.5''$. One of these outflows shows a
collimation degree of 10, the highest so far observed in massive
star-forming regions, and as high as the highest values known for
low-mass star formation. The observations indicate that high-mass
regions are more complex due to the clustered mode of formation, but
that the physical processes taking place are similar to those of their
low-mass counterparts.

There do exist other interferometric observations of high-mass
outflows, e.g., IRAS\,20126+2104 \citep{cesaroni 1997,shepherd 2000},
G192.16 \citep{shepherd 1998}, G35.2 \citep{gibb 2003}, but they are
either covering only the very inner region not tracing the large scale
outflow, or the spatial resolution does not exceed $5''$. 

In order to base the results obtained so far on a larger statistical
base and observe more massive outflows at adequate high spatial
resolution, we perform a series of high-resolution observations using
the Plateau de Bure Interferometer (PdBI) and BIMA. Here, we present
the results of PdBI observations of the outflow system in
IRAS\,19410+2336 which is part of a sample of 69 high-mass
protostellar objects (HMPOs) studied over recent years in great detail
\citep{sridha,beuther 2002a,beuther 2002b,beuther 2002c}.  At the very
center of the southern core, we detect H$_2$O and CH$_3$OH maser
emission as well as a weak 1\,mJy unresolved cm continuum source at a
spatial resolution of $0.7''$ (see Fig. \ref{single-dish}). Assuming
the latter to be due to optically thin free-free emission at 3.6\,cm,
it indicates a recently ignited but not very evolved massive object at
the cluster center \citep{beuther 2002c,sridha}.  It is also the most
deeply embedded massive star-forming region which was ever observed
with the {\it CHANDRA} X-ray satellite \citep{beuther 2002e}. So far,
we only knew the kinematic distances which suffer from a distance
ambiguity \citep{sridha}, but based on the new MSX and 2MASS data of
the region we believe the near kinematic distance to be more likely
(Bontemps, priv. comm.). Further on, we assume IRAS\,19410+2336 to be
at a distance of $\sim 2.1$\,kpc. The approximate bolometric
luminosity of the region is $10^4$\,L$_{\odot}$.  Single-dish 1.2\,mm
continuum observation reveal two massive adjacent dust cores, each
associated with a bipolar molecular outflow as observed in CO(2--1)
with $11''$ resolution (Fig.
\ref{single-dish}, \citealt{beuther 2002a,beuther 2002b}). The
single-dish data suggest a simple morphology in both cores which makes
the region a good candidate for further studies of collimated
outflows. The PdBI observations will show that the single-dish
observations provide only a simplified picture of the real outflow
structure.


\section{Observations}

\subsection{Plateau de Bure Interferometer (PdBI)}

We observed IRAS 19410+2336 in Summer 2001 with the Plateau de Bure
Interferometer at 2.6\,mm in the D (with 4 antennas) and C (with 5
antennas) configuration \citep{guilloteau 1992}. The simultaneously
observed 1\,mm data were only used for phase corrections because of
the poor Summer weather conditions. The 3\,mm receivers were tuned to
115.27\,GHz (USB) (centered at the $^{12}$CO $1\to0$ line) with a
sideband rejection of about 5\,dB. At this frequency, the typical SSB
system temperature is 300 to 400\,K, and the phase noise was below
30$^{\circ}$. The frequency resolution at 115\,GHz was
0.1\,km\,s$^{-1}$, but we smoothed the data to 1\,km\,s$^{-1}$ which
is sufficient to map the broad line wings. Atmospheric phase
correction based on the 1.3\,mm total power was applied. For continuum
measurements, we placed two 320\,MHz correlator units in the band to
cover the largest possible bandwidths. Temporal fluctuations of the
amplitude and phase were calibrated with frequent observations of the
quasars 1923+210 and 2023+336. The amplitude scale was derived from
measurements of MWC349 and CRL618, and we estimate the final flux
density accuracy to be $\sim 15\%$. To cover both cores a mosaic of 10
fields was observed. We obtain a $3\sigma$ continuum rms of $\sim
5$\,mJy. The continuum data are already presented by \citet{beuther 2002e}.

\subsection{Short spacings with the IRAM 30\,m telescope}

To account for the missing short spacings and to recover the line-flux,
we also observed the source in CO(1--0) with the IRAM 30\,m telescope
in Summer 2002. The observations were done remotely from Bonn in the
on-the-fly mode. Due to previous software changes the online velocity
correction was missing a term corresponding to some part of the earth
rotation.  This error was corrected offline by the AoD Gabriel
Paubert. The velocity is estimated to be correct within
0.1\,km\,s$^{-1}$.

The algorithm to derive visibilities from the single-dish data
corresponding to each pointing center is described by \citet{gueth
1996}. The single-dish and interferometer visibilities are
subsequently processed together. Relative weighting has been chosen to
minimize the negative side-lobes in the resulting dirty beam while
keeping the highest angular resolution possible. Images were produced
using natural weighting, then a CLEAN-based deconvolution of the
mosaic was performed. The final beam size of the 2.6\,mm data is
$3.9'' \times 3.6''$ (P.A. $33^{\circ}$).

\subsection{Calar Alto}

The near-infrared camera Omega Prime on the 3.5 m telescope on Calar
Alto/Spain was used to obtain 2.12$\mu$m narrow band H$_2$ and
$K^\prime$ wide field images of the region around IRAS\,19410+2336 on
June 11 2001. At a pixel scale of $0.4''$\,pixel$^{-1}$, the
1024$\times$1024 pixel array provides a field-of-view of $\sim$
6\farcs8 $\times$ 6\farcs8. The basic data reduction has already been
described in \citet{beuther 2002e}.  The narrow-band line mosaic and
the $K^\prime$ mosaic were then carefully registered. The Point Spread
Function (PSF) of the H$_2$ image was degraded to match the PSF of
the $K^\prime$ image with the poorer seeing; finally, the continuum
image was scaled such that stars had the same flux as in the narrow
band image.  This image was then subtracted from the (PSF-degraded)
narrow band image in order to provide the continuum subtracted narrow
band image (see Figure \ref{data} for the $K^\prime$ and H$_2$ image).

\section{Results}

\subsection{Millimeter continuum sources}
\label{mm-continuum}

\begin{figure}[ht]
  \includegraphics[angle=-90,width=8.7cm]{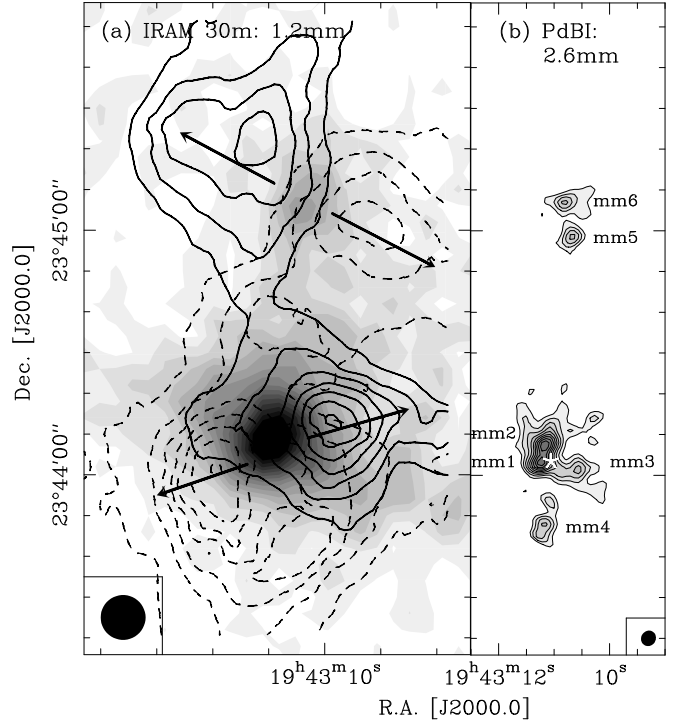}
\caption{The {\bf left} panel presents the single-dish observations 
already published in \citet{beuther 2002b}. In grey-scale we show the
1.2\,mm dust emission, and the solid and dashed contours outline the
blue and red CO(2--1) outflow emission. Both datasets were observed
with the IRAM 30\,m telescope at a spatial resolution of $11''$
(bottom left). The arrows mark the two outflows identified previously
from these observations. In the {\bf right} panel, we present the PdBI
2.6\,mm continuum data and label the sub-sources. The beam is
presented at the bottom right and the white cross marks the position
of the cm continuum source.}
\label{single-dish}
\end{figure}

The large-scale southern single-dish core splits up into at least
four sub-cores in the 2.6\,mm continuum (Fig. \ref{single-dish}),
whereas the northern core is resolved into 2 sub-sources in the
higher-resolution data. Assuming that the 2.6\,mm continuum is
produced by optically thin thermal dust emission, we can calculate the
masses and the column densities of the different sub-cores following
\citet{hildebrand 1983}:
\begin{eqnarray*}
M_{\rm{H_2}} & = & \frac{1.3\times10^{-3}}{J_\nu(T)}
\frac{a}{0.1\mu \rm{m}} \frac{\rho}{3 \rm{g cm}^{-3}} \frac{R}{100} 
\frac{F_{\nu}}{\rm{Jy}}\\
             &   & \times \left(\frac{D}{\rm{kpc}}\right)^2 
      \left(\frac{\nu}{2.4 \rm{THz}}\right)^{-3-\beta}\ [\rm{M}_{\odot}]\\
N_{\rm{H_2}} & = & \frac{7.8\times10^{10}}{J_\nu(T)\Omega} 
\frac{a}{0.1\mu \rm{m}} \frac{\rho}{3 \rm{g cm}^{-3}} \frac{R}{100} 
\frac{F_{\nu}}{\rm{Jy}}\\
             &   & \times \left(\frac{\nu}{2.4 \rm{THz}}\right)^{-3-\beta} 
    \ [\rm{cm}^{-2}]
\end{eqnarray*}
where $J_\nu(T)=[\exp(h\nu/kT)-1]^{-1}$ and $\Omega, a, \rho$, $R$,
and $\beta$ are the beam solid angle, grain size, grain mass density,
gas-to-dust ratio, and grain emissivity index for which we use the
values $3.9''\times 3.6''$ in radians, 0.1 $\mu m$, 3 g~cm$^{-3}$,
100, and 2, respectively \citep{hunter 2000}. Based on IRAS
far-infrared data, the dust temperature $T$ is assumed to be around
30\,K. As discussed by \citet{beuther 2002a}, higher dust opacity
indices (corresponding to a lower $\beta$) would result in lower
masses and column densities whereas lower temperatures would increase
the derived parameters. We estimate the results to be correct within a
factor 5.

The masses range between 40 and 180\,M$_{\odot}$ and the column
densities are in the $10^{24}$\,cm$^{-3}$ regime, corresponding to a
visual extinction of $A_{\rm{v}}\sim 1000$ (see Table
\ref{masses}). The mm sources mm1 and mm2 are of similar mass but the 
most prominent sub-core is mm1, which coincides with the H$_2$O and
CH$_3$OH masers and also with the weak cm continuum source
\citep{beuther 2002e}. The continuum 3$\sigma$ rms is $\sim$5\,mJy
equal to a mass sensitivity $>$20\,M$_{\odot}$ at a temperature of
30\,K. Assuming an $S_{\nu}\propto \nu ^4$ relationship (i.e., $\beta
= 2$) for the continuum dust emission in the mm regime, we compare the
single-dish flux at 1.2\,mm with the PdBI data at 2.6\,mm and find
that approximately 60\% of the total continuum flux is filtered out by
the interferometric observations. This most likely corresponds to
emission from the large-scale envelope of the core, whereas the PdBI
detected mm emission traces the very dense and compact
condensations. As the weak 1\,mJy cm continuum emission is most likely
due to optically thin free-free emission, its contribution in the
mm-regime is neglectable.

\begin{table*}[ht]
\begin{center}
\begin{tabular}{lrrrrrr}
\hline \hline
Source & R.A. & Dec & Flux & Peak & M & $N_{\rm{H_2}}$\\
       & [J2000.0] & [J2000.0] & [mJy] & [mJy/beam] & [M$_{\odot}$] & [$10^{24}$\,cm$^{-2}$] \\
\hline
mm1 & 19:43:11.24 & 23:44:03.4 & 29 & 19 & 132 & 4 \\
mm2 & 19:43:11.17 & 23:44:07.4 & 39 & 19 & 177 & 4 \\
mm3 & 19:43:10.59 & 23:44:01.5 & 14 & 9  & 64  & 2 \\
mm4 & 19:43:11.17 & 23:43:48.0 & 13 & 8  & 59  & 2 \\
mm5 & 19:43:10.68 & 23:44:58.4 & 10 & 8  & 45  & 2 \\
mm6 & 19:43:10.81 & 23:45:07.0 & 13 & 8  & 59  & 2 \\
\hline \hline			  
\end{tabular}
\caption{The mm sub-sources: positions, peak fluxes, integrated fluxes, 
derived core masses and column densities.}
\end{center}
\label{masses}
\end{table*}

\subsection{Outflow morphology}

Figure \ref{channel} shows a channel map of the merged PdBI+30\,m CO
emission in IRAS\,19410+2336, and we find a multitude of different
features. Based on the channel maps the velocity-regimes of outflowing
gas are defined as: blue wing [5;18]\,km/s and red wing
[26;47]\,km/s. The velocity spread does not vary significantly for the
different sub-outflows described below.

\begin{figure*}[ht]
\caption{Channel map of the merged PdBI and 30\,m CO(1--0) 
observations. The data are binned into 3\,km/s channels and labeled
each at the top-left. The arrows, ellipses, letters and markers
outline the outflows and mm sources (triangles) as discussed in the
text.}
\label{channel}
\end{figure*}

Comparing the single-dish CO emission presented in Figure
\ref{single-dish} with the high-resolution outflow map of the PdBI
(Fig. \ref{data} (a) and (b)) we see that the overall structures in
both datasets are similar, we just observe~-- as expected~-- much more
structure in the interferometric data. Identifying the same structures
makes us confident that the data reduction and merging process has
been done properly, but the sub-structures demand a different
morphological interpretation of the molecular outflows. Additionally,
the near-infrared shocked H$_2$ data show many emission knots
associated with the star-forming region (Fig. \ref{data}). Most of the
H$_2$ features can be correlated with the molecular gas, but for some
H$_2$ knots this is less obvious.  

We identify four molecular outflows in the southern core and three
outflows associated with the northern core. The identifications of the
outflows are not unambiguous, and we are still resolution
limited. Therefore, we note that the identifications below are just
one interpretation, others might be possible and only
very-high-spatial-resolution observations anticipated for next winter
will clarify the picture.

\begin{figure*}[htb]
\begin{center}
\end{center}
\caption{{\bf Top:} The left and middle panel show the blue 
(v[5;18]\,km/s) and red (v[26;47]\,km/s) CO emission separately
(merged PdBI and 30\,m data). The right panel then presents the
combined outflow emission and marks the outflows (arrows, ellipses and
letters), mm sources (triangles) and X-ray sources (stars) as
discussed in the text. {\bf Bottom:} The left and middle panels
show the $K^\prime$ and H$_2$ images from Calar Alto, respectively. The
labels in the middle panel mark the H$_2$ features discussed in the
text. The contours in the right panel again show the PdBI CO emission,
this time as an overlay on the H$_2$ map. The markers and errors are
the same as in the top panel.}
\label{data}
\end{figure*}


\subsubsection{The southern region}

The morphological interpretation of the four southern outflows is
sketched in Figures \ref{channel} and \ref{data}.

{\it South (A):} The structures labeled (A) and (B) could also be part
of only one outflow, but combining the CO and H$_2$ observations this
does not seem likely and we prefer the interpretation of two different
overlapping outflows. The outflow South (A) corresponds to the main
large scale outflow already identified in the single-dish observations
(see Figures \ref{single-dish}, \ref{channel} \& \ref{data}). It
runs in east-west direction and is centered approximately on the main
southern sub-core mm1 which also houses the cm-source and the maser
emission. Morphologically, one of the X-ray sources could also be the
center of the outflow, but the main mm source seems to be more
likely. 

{\it South (B):} The H$_2$ features (1, 4, 5, 6) and the corresponding
CO emission outline the ellipsoidal structure of outflow South (B)
sketched in Figures \ref{channel} and \ref{data}.
As the molecular emission at the eastern end of the outflow has a
strong gradient and stops rather instantaneously directly before
the H$_2$ feature (1), it is likely that both~-- CO and H$_2$~--
outline a bow-shock-like working surface between an underlying jet and
the surrounding molecular gas. The mm source mm3 likely houses the
driving source of outflow South (B). As discussed below, the H$_2$
features (4), (5) and (6) might outline the interface between the
outflow cavities of outflow (B) and the northern outflow (F).

{\it South (C):} The third southern outflow (C) is virtually
undetectable in the single-dish data because it is very small in
extent but clearly resolved in east-west direction in the
high-resolution PdBI data. The elongated H$_2$ emission feature (2) is
strongly associated with the blue molecular gas and outlines the same
morphological structure, indicating a highly collimated outflow.  The
H$_2$ knot (3) is at the eastern end of the red CO gas and can be part
of this outflow as well. South (C) is near the center of the region
and might also be driven by the mm source mm1 or another source
nearby. With the current spatial resolution it is likely that the
outflows South (A) and (C) emanate from the same mm continuum
condensation. At a distance of 2.1\,kpc, the beam-size of
$3.9''\times3.6''$ corresponds to a linear resolution of approximately
7900\,AU. Thus, it is well possible that a multi-component system is
evolving within mm1 and the outflows emanate from different members of
the system similar to observations in, e.g., HH288 \citep{gueth 2001}
or IRAS\,05358+3543 \citep{beuther 2002d}.

{\it South (D):} The fourth southern outflow labeled (D) is further to
the west and runs roughly north-south. It would already have been
possible to identify this flow in the single-dish data, but this
structure does not stand out prominently there
(Fig. \ref{single-dish}). We do not find a mm continuum or X-ray
source at the center of this outflow (Figs. \ref{channel} \&
\ref{data}). The driving source can well be hidden in the noise
because the continuum mass sensitivity is $>20$\,M$_{\odot}$ (\S
\ref{mm-continuum}). The H$_2$ feature (3) is in the vicinity of this
outflow but we cannot unambiguously attribute it to South (C) or (D).

\subsubsection{The northern region}

The northern region has a similarly complex morphological structure
and we suggest that at least three outflows are present there.

{\it North (E):} As for the outflows South (A) and (B), the northern
outflows (E) and (F) could also be parts of only one outflow system,
but for those structures, especially the CO emission indicates the
existence of two overlapping outflows. Outflow North (E) runs in a
collimated way roughly in east-west direction and corresponds to the
previously detected single-dish outflow in this region. Likely, the mm
source mm5 is the driving source of this outflow.

{\it North (F):} While the single-dish data suggested a simple
morphology (Fig. \ref{single-dish}), the red emission between the
northern and southern core stands out far more prominently in the PdBI
data. Therefore, we suggest that the red emission outlines the
ellipsoidal walls of a cavity-like outflow, as outlined in
Figures \ref{channel} \& \ref{data}. The blue-shifted emission to
the north-east shows a cavity-like structure as well and supports this
interpretation. In Figure \ref{channel}, the cavity structures are
best observed at velocities 18, 30, 33\,km/s. In this picture, the
H$_2$ features (4), (5) and (6) indicate shocked gas at the
interface between the cavities of the outflows (B) and (F).  There is
also a faint H$_2$ emission patch (12) in the north which could be
part of the northern cavity and which is just outside our observed
PdBI mosaic. The morphology suggests that the driving source of
outflow (F) is located within the mm source mm5 but it could also
be driven by mm6.

{\it North (G):} The finger-like blue filament pointing north has a a
red counterpart pointing south, both centering around the mm source
mm6. This morphology suggests a seventh outflow (G) running
north-south. Source mm6 is approximately at the center of this
structure and the only mm source where we find high-velocity gas as
well as H$_2$ emission (10) centered on (Figures \ref{channel} \&
\ref{data}). This suggest that the H$_2$ feature is directly produced
from the underlying high-velocity jet. At the northern tip, we find
weak H$_2$ emission (11) which can be associated with the blue CO
emission. Due to the spatial overlap with other outflows, it is not
clear how far south this North (G) proceeds and which part of the
emission belongs to North (G) or rather to South (A), (B) and (C). 
There are more H$_2$ emission knots (7, 8, 9) associated with the
southern part of this structure, but those features are difficult to
interpret and it is not clear whether they are part of North (G) or
belong to a completely different structure as outlined below.

\subsubsection{Further tentative outflow structures}

In the interface region between the northern and the southern core, we
find red/blue CO emission and the H$_2$ features (7), (8), and (9).

While it is possible that (7) and (8) are produced by outflow North
(G), the elongated H$_2$ emission (7) coincides with elongated red CO
emission. Additionally, Figure \ref{data} shows blue CO emission at the
H$_2$ knot (7). These observations allow the tentative detection of
another outflow (H) outlined by the dashed line in Figure
\ref{data}. We do not detect a mm continuum source
there as the outflow driver, but, as already mentioned, the continuum
sensitivity corresponds only to a mass sensitivity of
$>20$\,M$_{\odot}$, and a low-mass source can easily be hidden
there.

In addition to the red CO emission corresponding to outflow North (F),
there is also strong blue CO emission at the interface of the northern
and the southern star-forming cores. It is not entirely obvious where
this structure belongs to, but its north-east to south-west elongation,
the associated red CO emission and the H$_2$ knot (9) suggest that
even a ninth outflow (J) might be depicted by these structures. There
is no mm source detected in that region as well. We stress that the
structures (H) and (J) are even more ambiguous than the other seven
outflows discussed before. Higher spatial resolution and sensitivity
are needed to derive a conclusive picture of the different outflow
phenomena.

\subsection{Position-velocity diagrams}

\begin{figure}[ht]
\includegraphics[bb= 38 243 569 736, angle=-90,width=4cm]{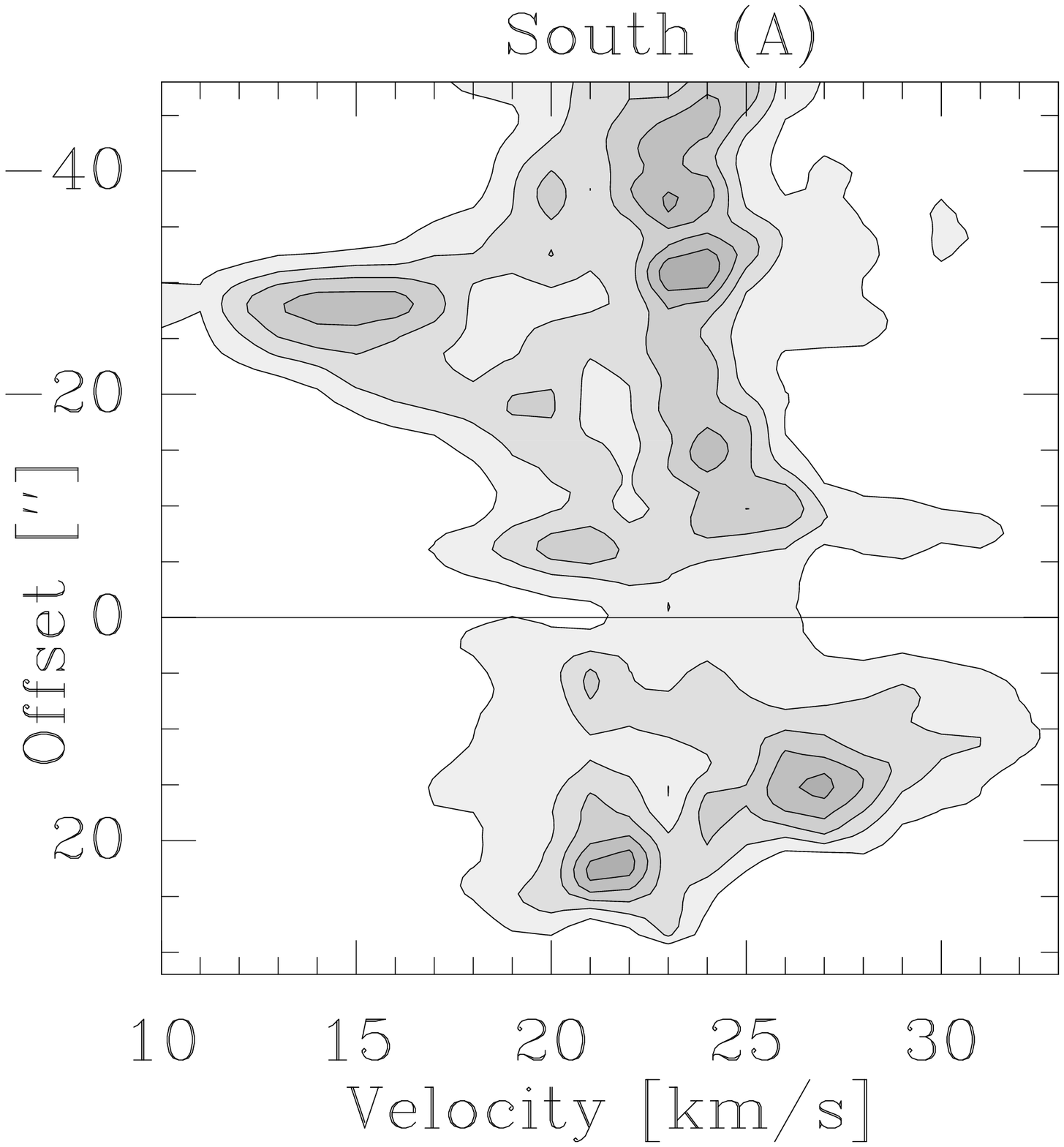}
\includegraphics[bb= 38 243 569 736, angle=-90,width=4cm]{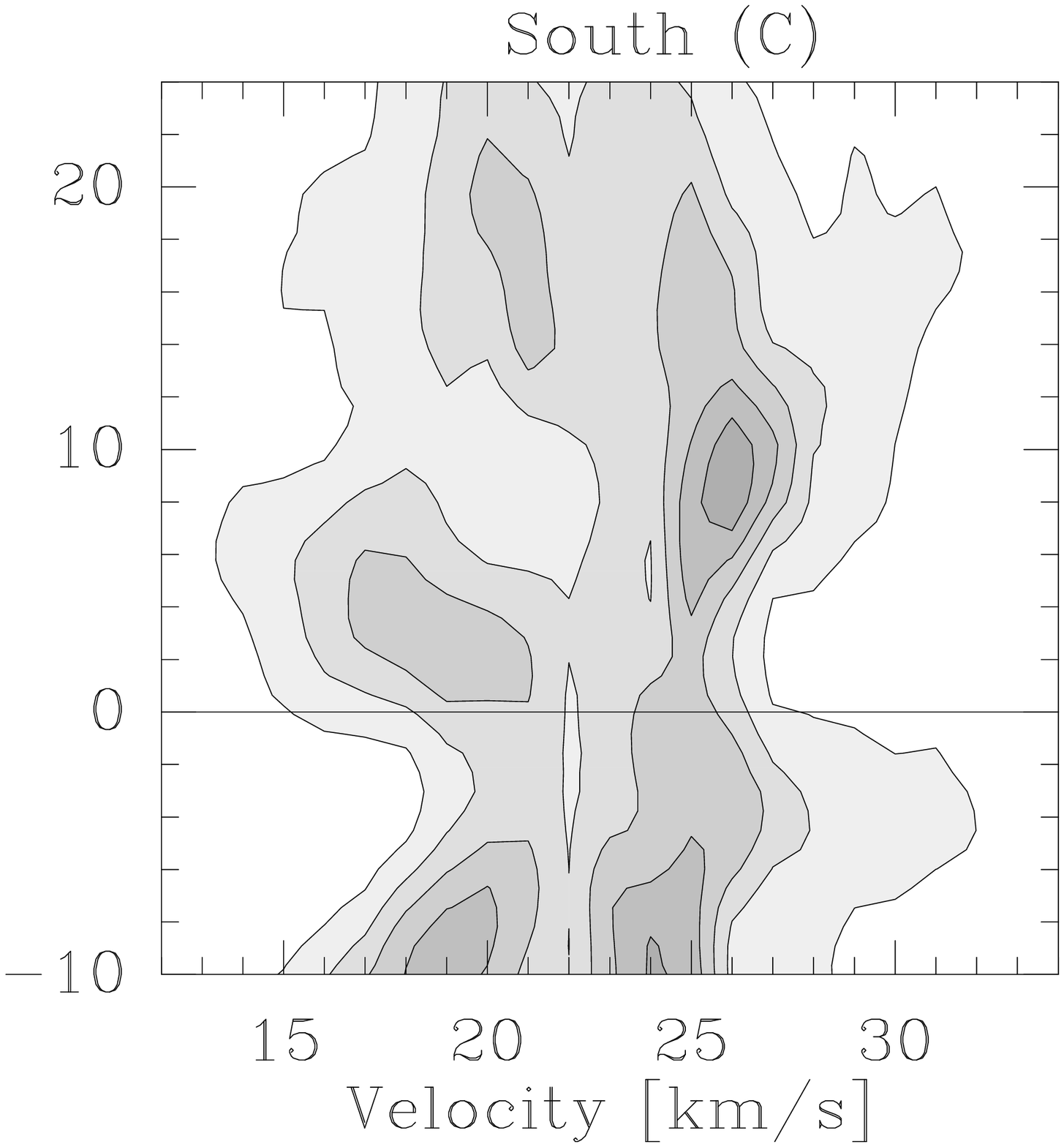}\\
\includegraphics[bb= 38 243 569 736, angle=-90,width=4cm]{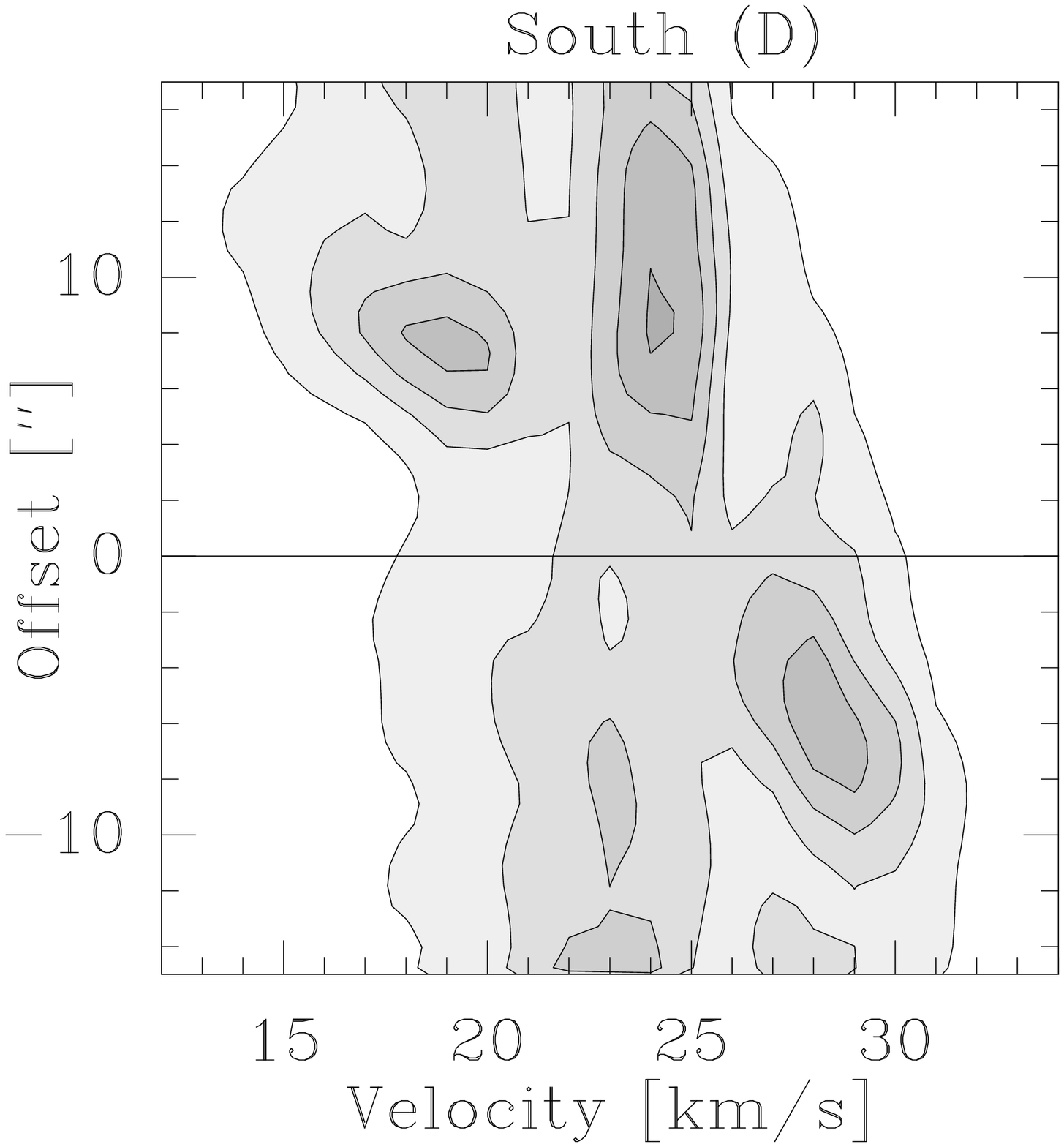}
\includegraphics[bb= 38 243 569 736, angle=-90,width=4cm]{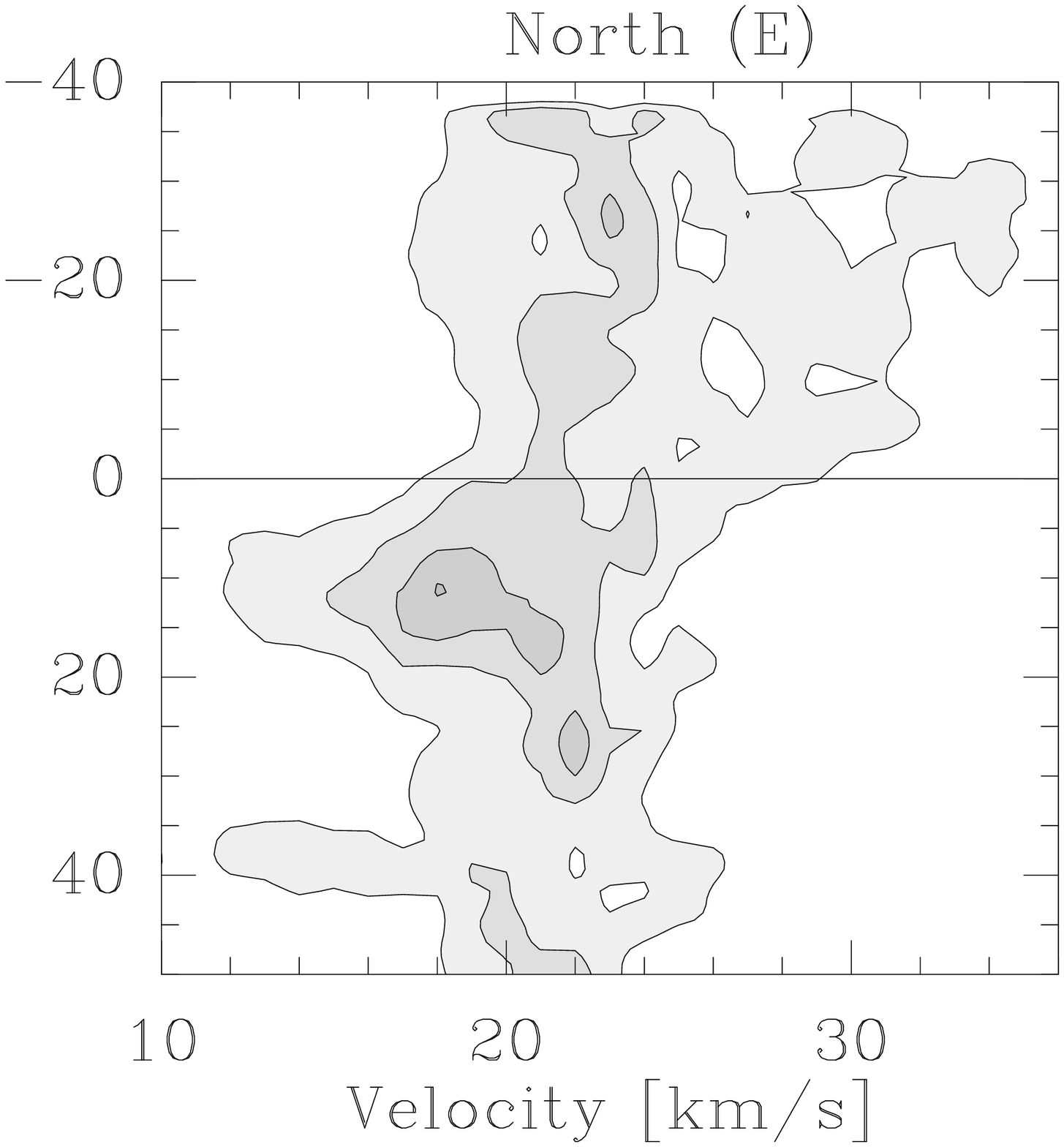}\\
\centerline{\includegraphics[bb= 38 243 569 736, angle=-90,width=4cm]{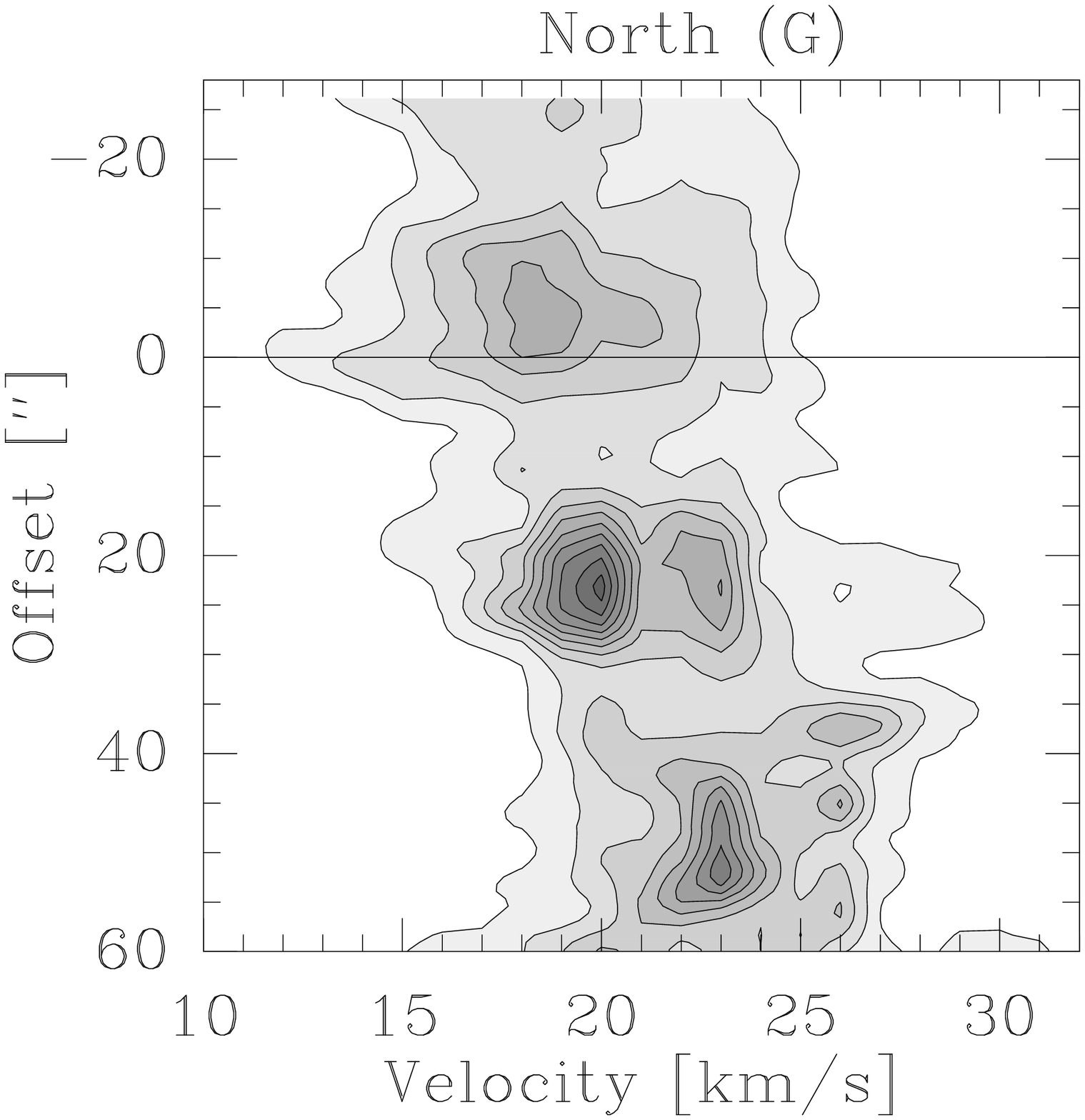}}
\includegraphics[bb= 38 243 569 736, angle=-90,width=4cm]{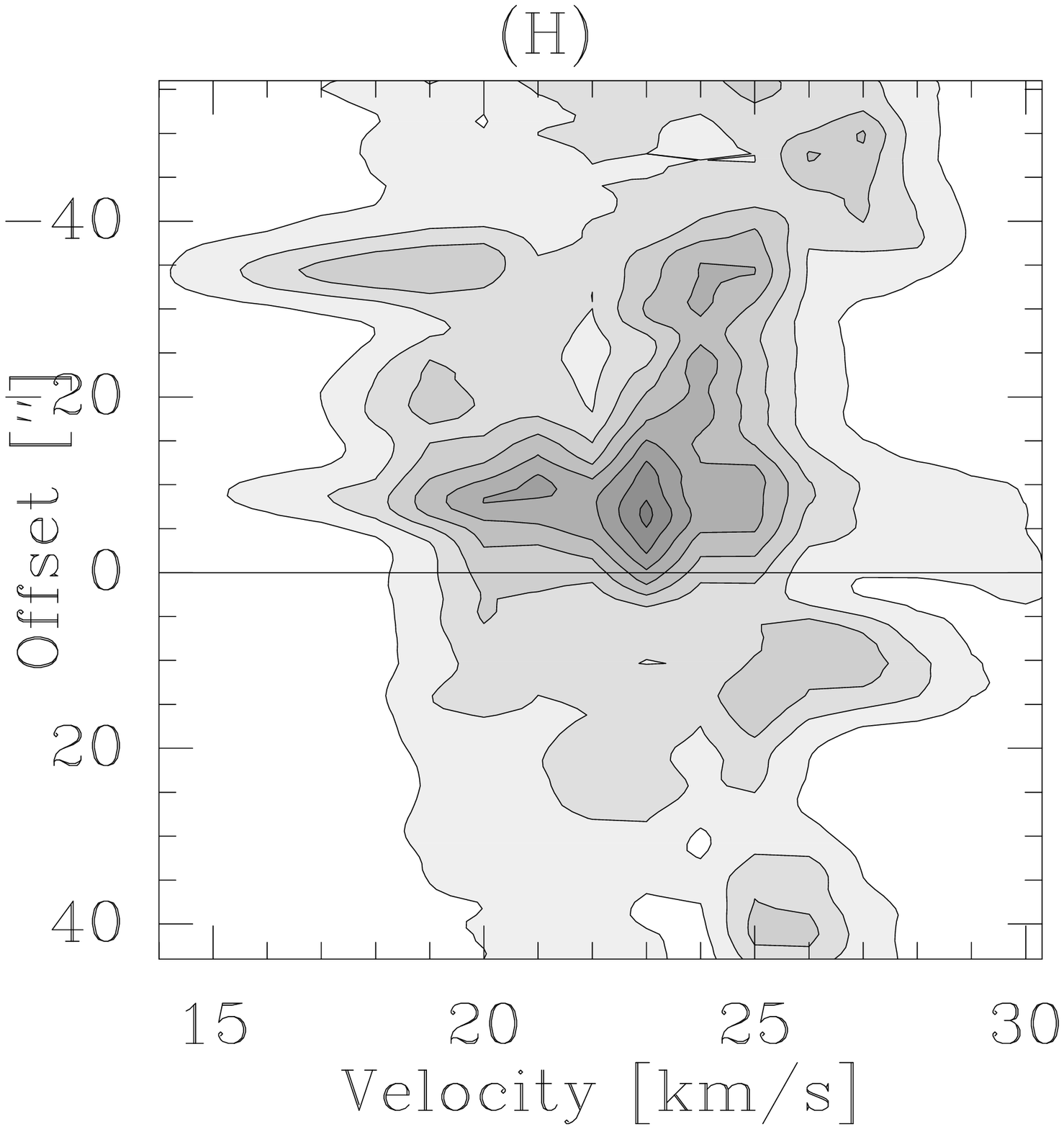}
\includegraphics[bb= 38 243 569 736, angle=-90,width=4cm]{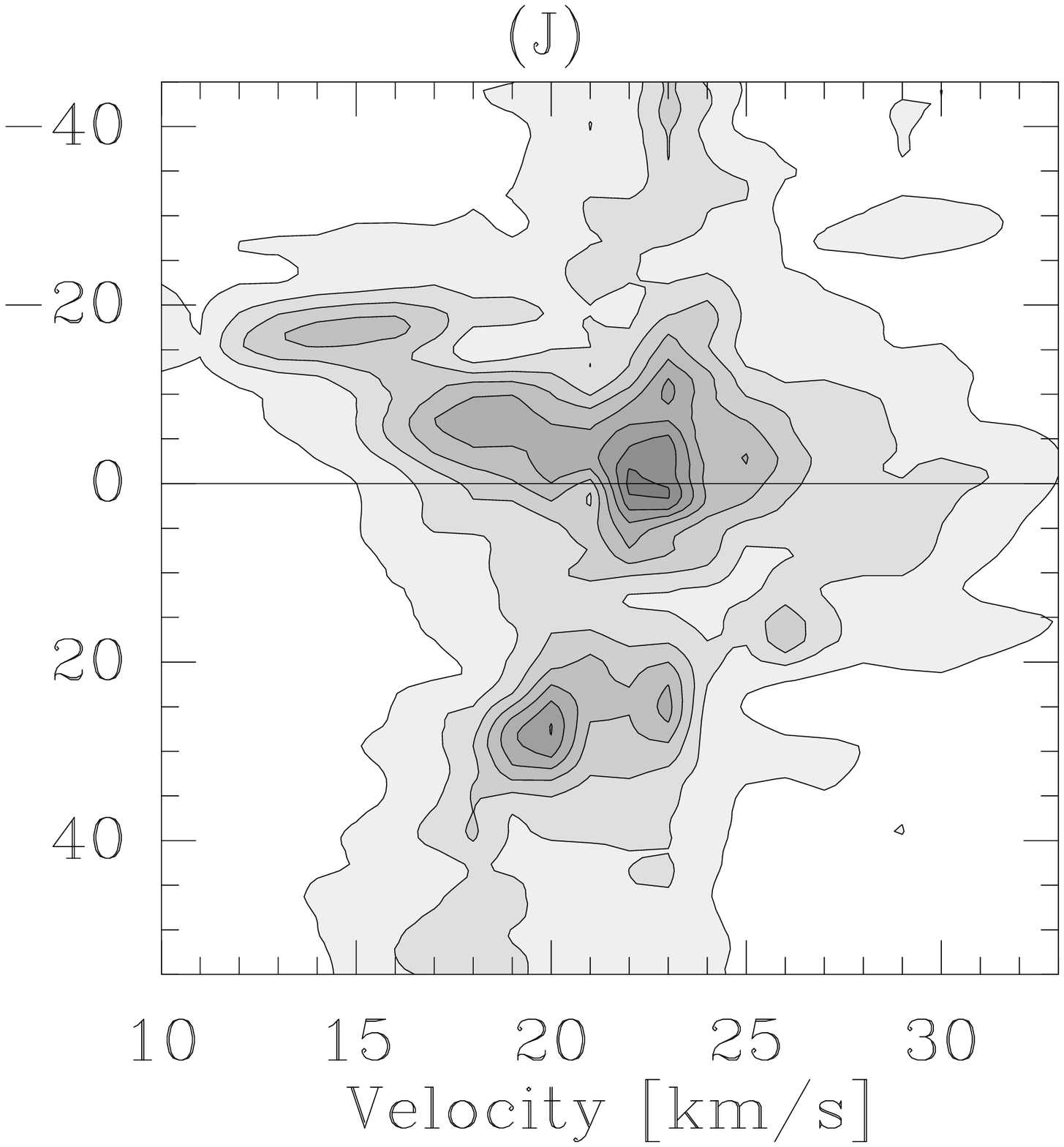}\\
\caption{Position-velocity diagrams for the five non-ellipsoidal outflows 
plus the two tentative ones (H) and (J). The horizontal line in each
diagram corresponds to the anticipated outflow center.The latter are
only assumptions for (H) and (J).}
\label{position-velocity}
\end{figure}

In spite of the seemingly chaotic morphology, the data still allow a
kinematic interpretation. Therefore, we present in
Fig. \ref{position-velocity} position-velocity diagrams for the five
non-ellipsoidal outflows (A, C, D, E, G) as well as for the tentative
structures (H) and (J). North (G) is the only outflow with
high-velocity gas at its center, for the other outflows, the
velocities increase with distance from the outflow center. This is
also the case for the position velocity diagrams corresponding to (H)
and (J), which is an additional indication that those features are
part of outflows. The position-velocity morphologies resemble a
Hubble-law, which can be interpreted due to the combination of
momentum conservation of the protostellar winds and density decreases
away from the driving sources of the outflows (e.g., \citealt{shu
1991}). Hydrodynamic jet-entrainment models also result in Hubble-like
position-velocity diagrams (e.g., \citealt{smith 1997,downes
1999}). The observed position-velocity diagrams resemble those known
from low-mass star formation. It has to be noted that already the
position-velocity diagrams derived from the single-dish data show
similar features.

\subsection{Outflow parameters}

In addition to morphological and kinematic interpretations, we
estimate masses, energetics and outflow rates of the individual
sub-flows. Because there is significant overlap between the various
outflows, often we cannot distinguish how much emission corresponds
to one or another of the overlapping outflows. Thus, some emission
features are used to derive the parameters for different
outflows. In the following, we present the outflow parameter for the
single-dish observations \citep{beuther 2002b}, compare them with the
outflow parameters derived for the same regions from the merged PdBI
and 30\,m data, and finally discuss the outflow characteristics of
the individual sub-outflow. It has to be noted that the single-dish
results presented here deviate from the parameters presented in Table
2 of \citet{beuther 2002b}, because in the former paper we erroneously
derived the mass in the red outflow wing of IRAS\,19410+2336 (south)
using a wrong source size. This error occured only in this case, and
the other parameters of Table 2 in \citet{beuther 2002b} are correct.

Opacity-corrected H$_2$ column densities $N_{\rm{b}}$ and $N_{\rm{r}}$
in both outflow lobes can be calculated by assuming a constant
$^{13}$CO/$^{12}$CO $1-0$ line wing ratio throughout the outflows
\citep{cabrit 1990}. \citet{choi 1993} found an average
$^{13}$CO/$^{12}$CO $2-1$ line wing ratio around 0.1 in 7 massive
star-forming regions, , corresponding to a $\tau
(^{13}$CO~$2-1)=0.1$. We adopt this value for IRAS\,19410+2336 as
well, and we assume 30\,K as average temperature in the outflow. The
mass of entrained gas M$_{\rm{out}}$, the dynamical timescale $t$ and
the outflow rate $\dot{M}_{\rm{outflow}}$ are calculated via:
\begin{eqnarray*}
M_{\rm{out}} &=& (N_{\rm{b}} \times \rm{size_b} + N_{\rm{r}} \times
\rm{size_r})\ m_{\rm{H_2}} \\ t &=&
\frac{r}{(v_{\rm{max}_b}+v_{\rm{max}_r})/2}\\
\dot{M}_{\rm{out}} &=& \frac{M_{\rm{out}}}{t} \\
\end{eqnarray*} 
with $\rm{size_b}$ and $\rm{size_r}$ the sizes of the outflow lobes,
$m_{\rm{H_2}}$, the mass of the H$_2$ molecule, and $v_{\rm{max}_b}$
and $v_{\rm{max}_r}$ the maximum velocities observed in each line
wing. A more detailed description how the outflow parameters are
determined is given in \citet{beuther 2002b}. According to
\citet{cabrit 1990} derived masses are accurate to a factor 2 to 4,
whereas the accuracy of dynamical parameters are lower, at about a
factor 10. 

\begin{table*}
\caption{Outflow results: masses $M_{\rm{b}}$ (blue), $M_{\rm{r}}$ (red) and $M_{\rm{out}}$ ($M_{\rm{out}}=M_{\rm{b}}+M_{\rm{r}}$) [$M_{\odot}$], momentum $p$ [$M_{\odot}$ km/s], energy $E$ [$10^{46}$erg], size [pc], time $t$ [$10^4$yr], mass entrainment rate $\dot{M}_{\rm{out}}$ [$10^{-4}$M$_{\odot}$/yr], mechanical force $F_{\rm{m}}$ [$10^{-3}$M$_{\odot}$km/s/yr] and mechanical luminsoity $L_{\rm{m}}$ [$L_{\odot}$] \label{output}}
\centering
\begin{tabular}{lrrrrrrrrrr}
\hline \hline
source & $M_{\rm{b}}$ & $M_{\rm{r}}$ & $M_{\rm{out}}$ & $p$ & $E$ & size & $t$ & $\dot{M}_{\rm{out}}$ & $F_{\rm{m}}$ & $L_{\rm{m}}$ \\ 
\hline
\multicolumn{11}{c}{Single-dish observations with the IRAM 30\,m in CO(2--1)}\\
\hline
South       &   11.2 &   19.8 &   31.0 &   683 & 1.5\,$10^{47}$ &  0.4 &  19510 & 1.6\,$10^{-3}$ & 3.5\,$10^{-2}$ &  63 \\ 
North       &   11.2 &   13.4 &   24.6 &   541 & 1.2\,$10^{47}$ &  0.5 &  20420 & 1.2\,$10^{-3}$ & 2.7\,$10^{-2}$ &  49 \\ 
\hline	     
\multicolumn{11}{c}{Merged PdBI and single-dish data of the single-dish regions in CO(1--0)}\\
\hline
South       &    8.2 &   10.4 &   18.6 &   651 & 2.6\,$10^{47}$ &  0.4 &  12790 & 1.5\,$10^{-3}$ & 5.1\,$10^{-2}$ & 167 \\ 
North       &    9.3 &   14.8 &   24.1 &   880 & 3.6\,$10^{47}$ &  0.5 &  13380 & 1.8\,$10^{-3}$ & 6.6\,$10^{-2}$ & 222 \\ 
\hline	     
\multicolumn{11}{c}{Merged PdBI and single-dish data of the individual outflows in CO(1--0)}\\
\hline
South (A)   &    2.6 &    3.0 &    5.6 &   192 & 7.6\,$10^{46}$ &  0.3 &   8330 & 6.7\,$10^{-4}$ & 2.3\,$10^{-2}$ &  75 \\ 
South (B)   &    1.7 &    1.6 &    3.3 &   107 & 4.1\,$10^{46}$ &  0.4 &  11300 & 2.9\,$10^{-4}$ & 9.5\,$10^{-3}$ &  30 \\ 
South (C)   &    0.6 &    0.7 &    1.3 &    46 & 1.8\,$10^{46}$ &  0.2 &   5060 & 2.6\,$10^{-4}$ & 9.0\,$10^{-3}$ &  29 \\ 
South (D)   &    1.6 &    2.6 &    4.2 &   153 & 6.3\,$10^{46}$ &  0.2 &   5950 & 7.1\,$10^{-4}$ & 2.6\,$10^{-2}$ &  87 \\ 
North (E)   &    3.5 &    4.2 &    7.7 &   265 & 1.1\,$10^{47}$ &  0.4 &  11300 & 6.8\,$10^{-4}$ & 2.3\,$10^{-2}$ &  76 \\ 
North (F)   &    4.5 &    8.1 &   12.6 &   472 & 2.0\,$10^{47}$ &  0.6 &  17840 & 7.1\,$10^{-4}$ & 2.6\,$10^{-2}$ &  91 \\ 
North (G)   &    1.9 &    6.8 &    8.7 &   359 & 1.6\,$10^{47}$ &  0.6 &  16950 & 5.1\,$10^{-4}$ & 2.1\,$10^{-2}$ &  77 \\ 
\hline \hline
\end{tabular}
\end{table*}

The first two sections of Table \ref{output} correspond to the
derived total outflow masses for the single-dish data and the merged
PdBI data, respectively. The chosen outflow region is based on the
single-dish data. Comparing the masses of entrained gas $M_{\rm{out}}$
from the single-dish and the PdBI observations, we find that the
merged interferometric data compare reasonably well with the
single-dish results. This approximate agreement is further support
that the merging of the interferometric and single-dish data, and also
their relative weighting, has worked properly. Additionally, it
indicates that single-dish observations are adequate to derive global
outflow properties.

More interesting are the parameters of all the sub-outflows derived
separately. We note that summing up the values from all southern and
northern outflows does not result in the total mass derived before,
because the morphologies vary too much: e.g., some emission features
believed from the single-dish data to be part of the southern flows
belong in fact to different flows of the northern
region. Furthermore, other emission patches are used to derive the
parameters for different outflows because we cannot always distinguish
unambiguously where they primarily belong to.

The derived outflow masses $M_{\rm{out}}$ are between 1\,M$_{\odot}$
for the smallest outflow South (C) and 12\,M$_{\odot}$ for North
(F). The outflow rates are high, of the order a few times
$10^{-4}$\,M$_{\odot}$\,yr$^{-1}$. Assuming momentum driven outflows,
a velocity ratio between the entrained gas and the underlying jet of
approximately 1/20 and a ratio between the jet-mass-loss rate and the
accretion rate of 0.3, such outflow rates result in accretion rate
estimates of the order $10^{-4}$\,M$_{\odot}$\,yr$^{-1}$ (for details
on the assumptions see \citealt{beuther 2002b}). Accretion rates of
that order are necessary to overcome the radiation pressure and form
massive stars via disk-mediated accretion processes
\citep{stahler 2000}. The accuracy of mass, age and other 
parameters is not good enough to compare individual properties between
the different flows in a meaningful way.

\section{Conclusions}
 
The new high-spatial resolution CO and H$_2$ observations of the
high-mass star-forming region IRAS\,19410+2336 resolve at least seven
(maybe even nine) bipolar outflows. The large number of outflows and
their interactions are an excellent example for the complexity in
high-mass star formation. The data show that the single-dish
observations simplify the morphological picture of the region, but
nevertheless give a good global picture of the region. Unfortunately,
we also find that even the present interferometric spatial resolution
is not sufficient to clearly disentangle the multiple overlapping
outflows, different interpretations for some of the features are well
possible.  CO and H$_2$ emission mostly trace different parts of the
same outflows, but it is not always possible to attribute all
observational features unambiguously within the same
interpretation. We note that here we present the~-- from our point of
view~-- most plausible interpretation of the outflow morphologies, but
it is well possible that follow-up observations will result in partly
different outflow identifications because of the complexity of the
region. Nevertheless, the whole picture is very suggestive of a
scenario, where the global outflow properties of massive star-forming
regions are a superposition of many different outflows emanating from
the evolving massive cluster.

The masses and energetics of each sub-flow are, as expected, below the
values derived from the single-dish data. Previous studies of other
massive molecular outflows have also shown multiple outflows but the
regions were dominated energetically by the most massive outflow
(e.g., IRAS\,05358+3543, \citealt{beuther 2002d}; IRAS\,23033+5951,
Wyrowski et al. in prep.). This is less clear in the case of
IRAS\,19410+2336: the northern core is dominated by outflow (F), but
it does not seem to be the case as much in the south, (A) contributes
only approximately 30\% of the mass estimated for the whole southern
core. The difference is lower if outflows (A) and (B) were not two,
but are different parts of only one flow. However, this question
cannot be resolved by the present observations. Combining the
observations from the literature with these new data, we find that
global core statistics based on single-dish observations are often
valid, but the averaging aspect of the single-dish data has to be kept
in mind.

Accretion rate estimates based on the outflow energetics are in the
$10^{-4}$\,M$_{\odot}$\,yr$^{-1}$ regime. Accretion rates of that
order are important to overcome the radiation pressure of the forming
massive stars, and thus build up the most massive objects via a
continuous disk-accretion process.

Of additional interest are the position-velocity diagrams, which
resemble a Hubble-like distribution as often reported in low-mass star
formation (e.g., \citealt{smith 1997,downes 1999}). This result is
different to findings based on CO single-dish observations from
\citet{ridge 2001} who report that they do not find Hubble-like
position-velocity dependence in any of their 11 studied massive
outflow sources. In contrast, resolving the two single-dish outflows
of IRAS\,19410+2336 into at least seven outflows with the PdBI
indicates that the kinematics of the individual outflows resemble the
features known from low-mass star formation. Furthermore, the H$_2$
emission associated with South (C) suggests the presence of outflows
in a well-collimated jet-like manner, as is found in low-mass
protostars. Together, this is further support to the hypothesis that
the physical processes in high-mass star formation are qualitatively
similar to those found in their low-mass counterparts, the main
difference being the clustered mode of formation and higher accretion
and outflow rates (e.g., \citealt{mckee 2002,maeder 2002,beuther
2002b,beuther 2002d}).

Concerning the dust continuum emission, in our previous single-dish
1.2\,mm observations with a spatial resolution of $11''$, we found
that the power-law density distribution does not continue to the very
center, and we derived an inner breakpoint at $5''$ for the southern
core from where on we modeled the density distribution as flat
\citep{beuther 2002a}. We suggested that this flattening is due
to fragmentation of the large-scale core into many sub-sources. The
new interferometric data directly confirm this hypothesis, and we
observe the fragmentation into sub-cores (Fig. \ref{single-dish}). The
spatial separation of, e.g., mm1 and mm2 is $\sim 5''$.

To summarize, the presented observations show the complex interaction
of several molecular outflows within the massive star-forming region
IRAS\,19410+2336. In spite of the complexity, the analysis of the data
supports the idea of massive star formation taking place in a
clustered mode by similar physical processes as known for low-mass
star formation, involving accretion through disks producing collimated
outflows.

Finally, we note that follow-up observations of this region are
necessary to understand the interplay of many molecular outflows
within one star-forming region more properly. In spite of the
similarity to low-mass star formation, the feed-back processes between
many protostars in a high-density environment should also produce
significant differences, IRAS\,19410+2336 can be used as a template
region for such studies. Regarding molecular outflows, different
molecular observations are needed, especially interesting is SiO which
is known to be a shock and jet tracer. Furthermore, we are still
resolution limited, and it will be crucial to re-observe the
region in CO(2--1) with significantly higher spatial resolution.

\begin{acknowledgements}
We like to thank the IRAM Grenoble staff for the help during
observations and data reduction, especially F. Gueth and
D. N\"urnberger. Additionally, we thank G. Paubert, the AoD at the
30\,m telescope, for correcting the velocity information of the
single-dish data offline. S. Bontemps helped resolving the distance
ambiguity. H.B. acknowledges financial support by the
Emmy-Noether-Programm of the Deutsche Forschungsgemeinschaft (DFG,
grant BE2578/1).
\end{acknowledgements}

\end{document}